\documentclass{ws-p9-75x6-50}

\begin{document}

\title{Complete null data for a black hole collision}

\author{Sascha Husa${}^{1,2}$,
        Jeffrey Winicour${}^{1,2}$}

\address{Department of Physics and Astronomy \\
         University of Pittsburgh, Pittsburgh, PA 15260\\ and\\
         Max-Planck-Institut f\" ur Gravitationsphysik,
         Albert-Einstein-Institut \\
         14476 Golm, Germany}

\author{Roberto Gomez}

\address{Department of Physics and Astronomy \\
         University of Pittsburgh, Pittsburgh, PA 15260}
\maketitle

\abstracts{
We discuss  a sequence of numerically constructed geometries describing
binary black hole event horizons -- providing the  necessary input for
characteristic evolution of the exterior spacetime. Our sequence approaches
a single Schwarzschild horizon as one limiting case and also includes
cases where the horizon's crossover surface is not hidden by a
marginally anti-trapped surface (MATS).}

In previous work, we presented a conformal horizon model of a binary black
hole which generates the ``pair-of-pants'' event horizon structure found in
the axisymmetric head-on collision of black holes~\cite{ndata}.
The conformal model constructs a null surface geometry by conformal rescaling
of the Minkowski lightcone of a spheroid of eccentricity $\epsilon$, and
supplies part of the data for a characteristic evolution
{\em backward in time} along ingoing null hypersurfaces. The strategy and details behind this new
approach to determine a part of the exterior space-time and the waveform
emitted in the post-merger phase has been outlined elsewhere
~\cite{ndata,asym,kyoto,gr-qc/0009092}.
Recently we generalized our previous study of the intrinsic geometry
of the horizon to a full stand-alone description of both the intrinsic and
extrinsic horizon data necessary to implement a characteristic evolution
\cite{gr-qc/0009092}.
In such an evolution of the exterior spacetime, along a
family of {\it outgoing} null hypersurfaces, we choose the inner boundary
as the null worldtube representing a {\em white hole horizon}.
This horizon pinches off in the future, where its
generators either caustic or cross each other (such as they do at the vertex of a null
cone). We work in the coordinate system
first introduced by Sachs to formulate the double-null
characteristic initial value problem~\cite{sachsdn}.

In the close approximation \cite{close} regime of a white hole fission, when
reinterpreted in the time reversed sense of a black hole merger, we find that the
individual black holes merge inside a white hole horizon corresponding to the
marginally anti-trapped branch of the $r=2M$ Schwarzschild surface. In
\cite{gr-qc/0009092} we demonstrate numerically that 
in the non-perturbative regime an entirely different
scenario is possible, in which the individual black holes form and merge without
the existence of a MATS on the event horizon.
Since the Bondi surface area coordinate is singular on a MATS, the 
absence of a MATS is required for a Bondi evolution backward in
time throughout the space-time region exterior to the black holes. 
In our present approach, we deal with the Bondi boundary  ${\cal B}$,
defined by $\partial_\lambda r=0$, rather than the MATS. We prove in 
\cite{gr-qc/0009092} that a marginally trapped surface cannot form before
the Bondi boundary on a white hole horizon, thus in the time reversed black
hole picture, absence of a Bondi boundary implies absence of a MATS.

Of special physical importance is the location of the {\em crossover
surface} ${\cal
X}$, where the horizon pinches off, relative to the surface ${\cal B}$. 
For small $\epsilon$ the fission (located on the equator of ${\cal X}$) is
``hidden'' beyond ${\cal B}$ in the sense that it is not visible to observers
at ${\cal I}^+$. 
{}From the view of a black hole merger, the individual black
holes would merge inside a white hole horizon.
Our analytical results described in \cite{gr-qc/0009092} suggest that
sufficient nonlinearity might cause the white hole fission to occur prior to
${\cal B}$.
This scenario can indeed be demonstrated by numerical integration of the
equations underlying the conformal horizon model. At an
early time, the equilibrium conditions on the white hole horizon imply that
$r=2M$ and $\partial_\lambda r= -u/4M >0$ ($u$ is a suitably normalized
affine coordinate at the horizon). As the horizon evolves, the surface
area $r$ decreases along all rays. The outward expansion measured by
$\Theta_{OUT}=2\partial_\lambda r /r$ also initially decreases along all rays,
although this process can be reversed by the growth of nonlinear terms, as
indicated by the ray-averaged behavior. In
the close approximation, the expansion goes to zero along all rays before the
horizon pinches off, i.e. the crotch at the center of the pair of pants is
hidden behind a MTS. The crucial question in the nonlinear regime is whether
the horizon can pinch off before the formation of a Bondi boundary, i.e.
the issue is who wins the race towards zero, the radius $r$ or the
expansion $\Theta_{OUT}$ along some ray.

We have conducted this race for a sequence of models in the range
 $0 \le \epsilon \le 10^{-2}$ and
monitored the minimum value of the expansion of the outgoing null rays on the
horizon over the sphere, and of the Bondi radius of the horizon.
We find that indeed for small $\epsilon$ we can confirm the close
limit picture where a Bondi boundary forms before the radius has
changed significantly. But for sufficiently large  eccentricity $(\epsilon=10^{-3})$,
the radius {\em does} win the race by a sudden plunge to zero before the
expansion has undergone any appreciable change. This is a genuine nonlinear
effect: While the expansion begins with a flying start (its initial slope)
and gets accelerated by linear effects, the radius starts from rest and
``only'' gets accelerated by quadratic or higher nonlinearities.

\section*{Acknowledgments}
This work has been partially supported by NSF grants PHY
9510895 and PHY 9800731 to the University of
Pittsburgh. R.G. thanks the Albert-Einstein-Institut for
hospitality. Computer time was provided by the
Pittsburgh Supercomputing Center and by NPACI.

\end{document}